# Resource Management Services for a Grid Analysis Environment

Arshad Ali[1], Ashiq Anjum[1,4], Tahir Azim[1], Julian Bunn[2], Atif Mehmood[1], Richard McClatchey[4], Harvey Newman[2], Waqas ur Rehman[1], Conrad Steenberg[2], Michael Thomas[2], Frank van Lingen[2], Ian Willers[3], Muhammad Adeel Zafar[1]

[1]*National University of Sciences & Technology, Rawalpindi, Pakistan*
*Email: {arshad.ali, ashiq.anjum, tahir, atif.mehmood, waqas.rehman, zafar.adeel}@niit.edu.pk*
[2]*California Institute of Technology, Pasadena, USA*
*Email: {Julian.Bunn, fvlingen}@caltech.edu, {newman, conrad, thomas}@hep.caltech.edu*
[3]*European Organization for Nuclear Research, Geneva, Switzerland*
*Email: Ian.Willers@cern.ch*
[4]*University of the West of England, Bristol, UK*
*Email: Richard.McClatchey@uwe.ac.uk*

## Abstract

*Selecting optimal resources for submitting jobs on a computational Grid or accessing data from a data grid is one of the most important tasks of any Grid middleware. Most modern Grid software today satisfies this responsibility and gives a best-effort performance to solve this problem. Almost all decisions regarding scheduling and data access are made by the software automatically, giving users little or no control over the entire process. To solve this problem, a more interactive set of services and middleware is desired that provides users more information about Grid weather, and gives them more control over the decision making process. This paper presents a set of services that have been developed to provide more interactive resource management capabilities within the Grid Analysis Environment (GAE) being developed collaboratively by Caltech, NUST and several other institutes. These include a steering service, a job monitoring service and an estimator service that have been designed and written using a common Grid-enabled Web Services framework named Clarens. The paper also presents a performance analysis of the developed services to show that they have indeed resulted in a more interactive and powerful system for user-centric Grid-enabled physics analysis.*

## 1. Introduction

Efficient resource management capabilities are one of the most important features in all Grid computing middleware. Resources may include processing facilities, storage elements, and network bandwidth that are required by a Grid-enabled application (also referred to as a job). Currently available Grid middleware carries out the task of resource management such as job scheduling over multiple computational facilities, selecting and accessing datasets from suitable storage elements, out of sight from the user. Although this feature is desirable in many respects, such as shielding users from the complexity of the underlying Grid system and allowing inexperienced users to harness the power of the grid, it also prevents advanced users from exploiting the maximum power offered by the computational grid. Advanced users prefer to control and steer their jobs more effectively by getting more information about the state of the Grid resources, observing the progress of their running jobs, and moving or restarting jobs on other resources that are more powerful or under utilized. By doing so, they can gain more control over the available resources to optimize (or improve) performance of those resources. This would also facilitate the development of more intelligent agents that could observe and learn from the actions of advances users, and work out improved optimization strategies for automated resource management activities.

In this paper, we discuss a set of services that have been developed to provide end users, in particular advanced end users, the ability to get more information about available resources and to gain more control over the execution of their jobs. These services have been written within the framework of a Grid Analysis Environment (GAE) [1]. Three main services are described:

- A steering service developed to provide dynamic and adaptive resource management due to the volatile nature of a Grid environment. This service also provides users information about the progress of their jobs and lets them interact with their jobs through actions like pausing or restarting jobs, or moving them to better resources.
- An estimator service that allows users and other services to estimate the cost of doing a number of tasks, such as running a job, and transferring a file.
- A job monitoring service that enables the steering service (as well as clients) to collect information about the status of running jobs and the resources being used by them.

This paper is organized as follows. In Section 2, we describe a brief background of the problem and the requirement of greater control over job execution on the Grid. In Section 3, we describe the overall architecture of the system and the mutual interaction between the web services through the Clarens web services framework [2]. We then go on to describe the implementation of the services in Sections 4, 5 and 6. We present some performance results in Section 7, before giving an overview of related work in Section 8 and concluding in Section 9.

## 2. Grid Usage in High Energy Physics

The CMS [3] experiment is one of the four High Energy Physics (HEP) experiments which are part of the LHC (Large Hadron Collider) program under construction at CERN, and will become operational in the year 2007. In addition to this construction, several physicists, engineers and computer scientists work together on a software infrastructure that will allow physicists to analyze data produced by the CMS detectors.

From a computing point of view, physics analysis is a very complex task since it includes several computing challenges. On the one hand, large amounts of data (of the order of Tera- and Peta-bytes) have to be stored and replicated to several geographically distributed sites. Next, physics analysis code needs to be in place that understands the physics process in the detector. Finally, a distributed software system is required to identify where the requested data is located, to determine the best and closest available locations for executing the physics analysis code, and finally to submit jobs for execution.

Current Grid tools used by high-energy physics are geared towards batch analysis. A large number of computing jobs are split up into a number of processing steps (arranged to follow a directed acyclic graph structure) and are executed in parallel on a computing farm. The only interaction between the user and the batch job is the ability to display the progress of the computing job and the ability to restart processing steps that may have failed due to error.

As stated earlier, advanced users might be able to make better decisions regarding which jobs to execute on high priority and what computational facilities to use to execute those jobs. The resource management services outlined in this paper provide users with this capability. While it is still possible for average users to continue to use the normal system of automatic job scheduling and progress monitoring, advanced users can get potentially better productivity by steering their jobs manually for faster performance.

## 3. Web Services Based Architecture

In order to provide advanced users the ability to get more productivity from the available Grid infrastructure, we have developed a collection of services that can interact with each other to share monitoring and resource information, store the state of users' analysis sessions, and allow users to make their own choices about job execution.

These services have been developed as part of the Grid Analysis Environment (GAE) [1]. The GAE envisages the development of an ensemble of web services cooperating to form a more interactive analysis environment, in order to meet the interactive analysis requirements specified by the Particle Physics Data Grid (PPDG) CS-11 working group [4]. The Clarens web service hosts are the backbone of this GAE. Clarens offers a web service framework for hosting the GAE web services, and provides a common set of services for authentication, access control, and for service lookup and discovery. Clarens enables users and services to dynamically discover other services and resources within the GAE through a peer-to-peer based lookup service [5].

In this paper, we describe a number of services that we have developed in order to meet the needs of advanced users described in Section (1) and (2) above.

These services include the steering service, the job monitoring service, and the estimator service. The steering service provides users with real-time control of their job submissions, allowing users to steer their jobs in order to achieve sustained performance. The purpose of the Job Monitoring service is to provide real-time job monitoring information and status feedback while operating in close interaction with an execution service (which can be based on any execution engine such as Condor [6]). The estimator service can be used to provide estimates of the resources required by a job, based upon historical information. It also provides information (static estimates) to the scheduler for scheduling decisions. A rough outline of the interaction between the services that make up this architecture is depicted in Figure 1.

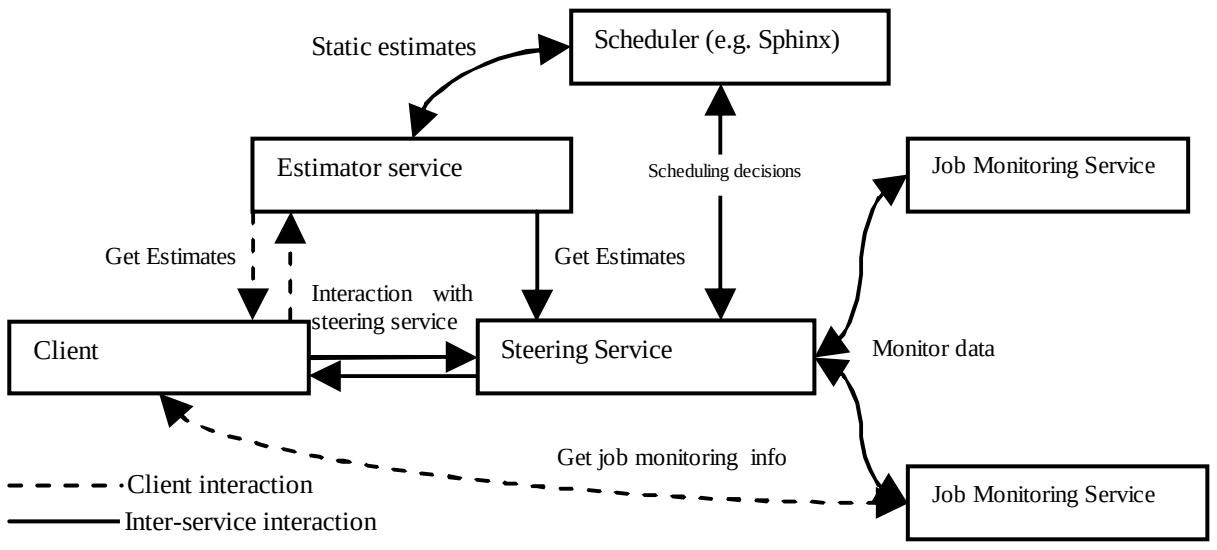

**Figure 1. Mutual interaction between the resource management services**

Clients can contact any of these services directly in order to get the information they require about their jobs or the state of available resources. In addition, the steering service uses information from both the job monitoring and the estimator service in order to make its decisions. The steering service uses information from the job monitoring service to determine the resources being used by various jobs and track their progress. Similarly, the steering service determines the estimated time to completion of a job or the transfer time of a file by invoking the estimator service. The job monitoring service uses data from the execution service to extract monitoring information related to a currently executing job. In this way, each of the services leverages the capabilities of the other services to share information and use that information to make better choices.

The services have been designed as SOAP/XMLRPC web services to ensure a modular architecture, to enable clients to access these services in a language-neutral manner, to enable the services to communicate with each other over a wide area or local area network, and to allow any future services to utilize the information published by these services. All the services discussed in this article have been deployed using the Java version of the Clarens web services framework.

## 4. Steering Service

The Steering Service is the component of the GAE architecture that allows users to interact with submitted jobs. The Steering Service provides constant feedback of the submitted jobs to the users. It also allows the users to change the status of their jobs. This includes kill, pause, and resume, change priority of the job or moving the job to some other execution site.

### 4.1. Architecture

The architecture of the Steering Service is shown in the following diagram. A detailed description of the various components follows in the next section.

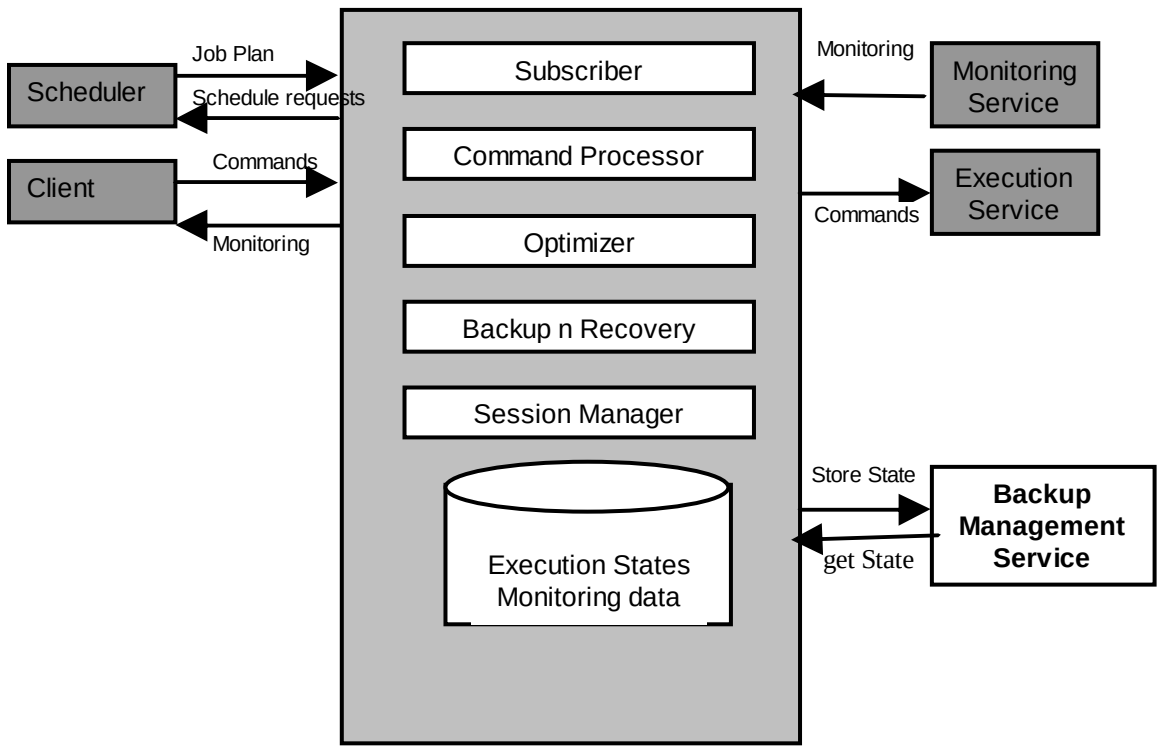

**Figure 2. Design of the Steering Service**

### 4.2. Components

The following components make up the Steering Service.

**4.2.1. Subscriber.** A scheduler (e.g. Sphinx [7] in GAE) sends a “concrete job plan” (a job plan precisely describing the nodes where the job will be executed) to the Steering Service. The Subscriber analyzes the received job plan to get the list of Execution Services to be used for the execution of the job for which the job plan has been sent.

**4.2.2. Command Processor.** The Command Processor handles the requests of the client and requests of the optimizer to perform job control e.g. kill, pause, resume, move job. Requests for job redirection are sent to the scheduler (Sphinx).

**4.2.2. Optimizer.** The optimizer contacts the Quota and Accounting Service (currently, just a trivial prototype) to find the cheapest site for job execution, and interacts with the Estimators to determine the site that can execute the task faster. Based on the information gathered, the job is redirected to the “Best Site”. The meaning of “Best Site” depends on the optimization preference chosen (cheap or fast execution). The expected execution time, calculated using the Estimator Service, includes the run time, queue time, and file transfer time estimates for job execution on a particular site.

**4.2.4. Backup and Recovery.** This module continuously checks all the Execution Services (on which the different tasks of a job are running) for failure. In case of the failure of the Execution Service, the Backup and Recovery module contacts Sphinx to allocate a new execution service. The scheduler will then resubmit the job on that new execution service.

If a running job fails, the Steering Service notifies the client about the failure. It then contacts the execution service to get all the local files that were produced by the failed job. For completed jobs, the Backup and Recovery module notifies the client about the completion of the job and gets the execution state from the execution service. This execution state is made available for download on the web interface.

**4.2.5 Session Manager.** This module makes sure that the authorized users steer the jobs.

## 5. Job Monitoring Service

The Job Monitoring Service provides the facility of monitoring jobs that have been submitted for execution, and provides the job monitoring information to the Steering Service. The Job Monitoring Service also provides an easy-to-use API for retrieval of job monitoring information such as job status, remaining time, elapsed time, estimated run time, queue position, priority, submission time, execution time, completion time, CPU time used, amount of input IO and output IO, owner name and environment variables.

The Job Monitoring Service also continuously monitors the jobs that have been submitted and sends an update to MonALISA [8] whenever the state of a job changes. The main components of the Job Monitoring Service are described below, and their interactions are shown in Figure 3.

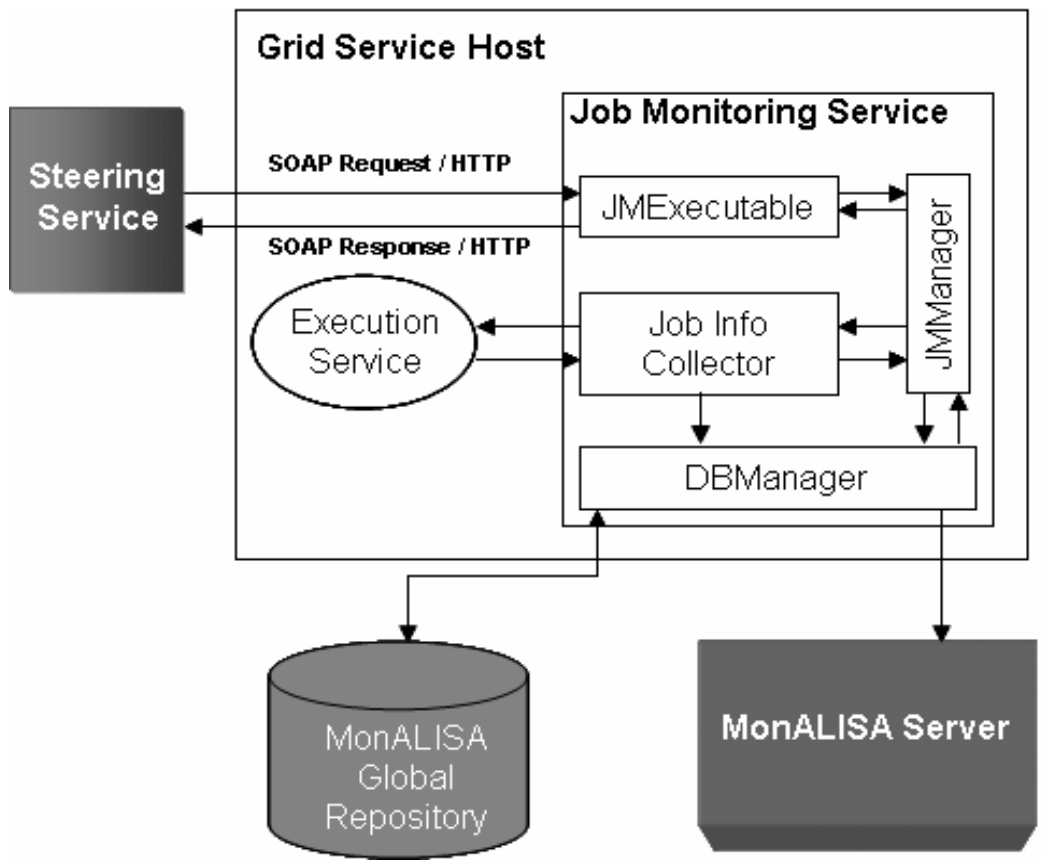

**Figure 3. Components of the Job Monitoring Service**

### 5.2. Job Information Collector

The role of the Job Information Collector module is to monitor the jobs that have been scheduled. The Job Information Collector interacts with the Execution Service to provide real-time job monitoring information. The Job Information Collector functions in two ways:

• It monitors the job execution and whenever the job is completed or terminated due to an error, it sends an update request to the DBManager for that job.

• It provides the monitoring information of the running jobs to the JMManager when requested.

### 5.3. JMExecutable & JMManager

The JMExecutable serves to forward requests by the Steering Service to the JMManager. The JMManager handles the flow of information within the Job Monitoring Service. The JMManager gets the monitoring information either from the DBManager or from the Job Information Collector. It first queries the DBManager and if the information is not found in its repository, the request is forwarded to the Job Information Collector. The information is then sent to the Steering Service via the JMExecutable.

### 5.4. DBManager

Each Job Monitoring Service instance has a database repository. The access to this repository is controlled by the DBManager. The DBManager publishes the job monitoring information to MonALISA.

## 6. Estimators

The Estimator Service (or simply the estimators) is used to predict the resource consumption of a job. The Estimators API provides the following estimators:

### 6.1. Runtime Estimator

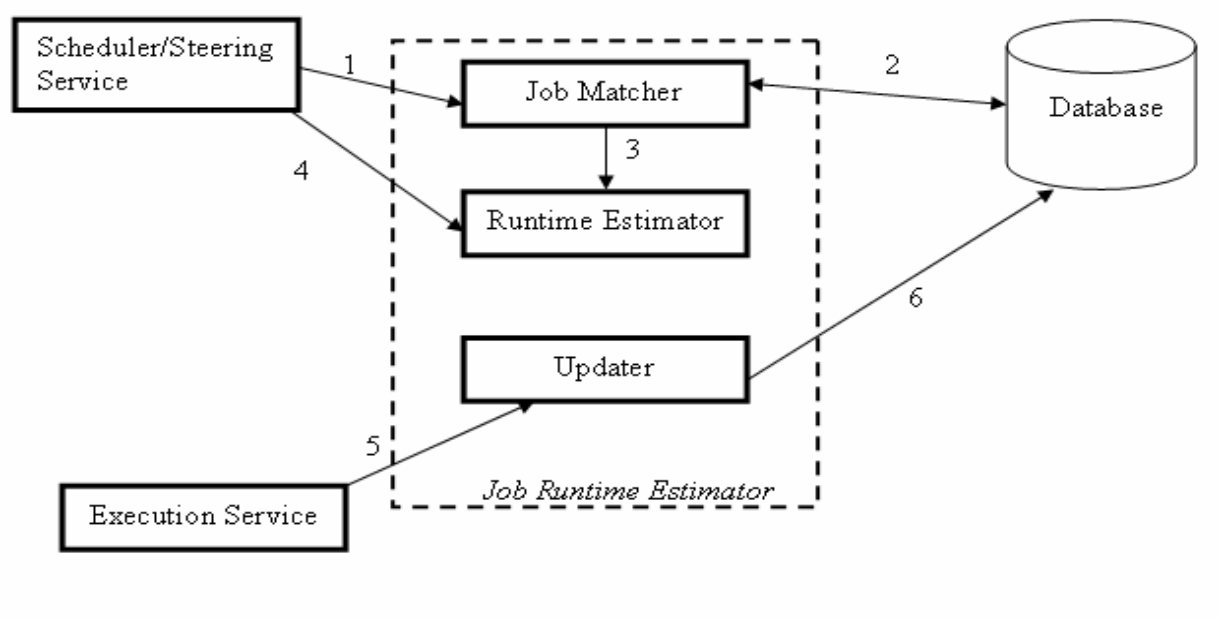

**Figure 4. Design of the Runtime Estimator**

The Runtime estimator computes the estimated runtime of an input task (the atomic component of a job) at a single execution site. Runtime is estimated by using a history-based approach. History based runtime prediction algorithms operate on the idea that tasks with similar characteristics generally have similar runtimes [9] (although this might not be the case for many kinds of tasks). We maintain a history of tasks that have executed along with their respective runtimes. To estimate the runtime, we identify similar tasks in the history and then compute a statistical estimate (the mean and linear regression) of their runtimes. We use this as the predicted runtime. A decentralized approach is used for history maintenance, so whenever a job is submitted to the scheduler, it performs the following sequence of events for each task in the job description file:

a. The scheduler contacts the available execution sites, and passes the task's attributes to the Execution service (at every execution site). Of course, this depends on the availability of the runtime estimator at each of the sites.

b. The execution service (which will be hosting the execution service) estimates the run time of the task using the estimator.

c. The estimated run time of step (b) will then be returned back to scheduler.

d. After getting estimates from every execution site, the scheduler will then contact the MonALISA repository to get the status of load at execution sites.

e. Based on the estimated run time and load status from step (b) and (d), the Scheduler will then select a site that has the least estimated run time and where the queue time for the task is a minimum.

For detailed information on how job matching is performed and estimated runtimes are computed, see [10].

### 6.2. Queue Time Estimator

The Queue Time Estimator estimates the time a task will spend in a queue waiting for its turn to start execution. This queue time is used by the Steering service to let the user know how much time his/her task will spend in queue.

In order to estimate the queue times the following sequence of events take place:

a. The Condor ID of the task is provided as the input to the Queue Time Estimator, and the Queue Time Estimator then contacts the execution service (e.g. Condor) and retrieves from the queue Condor IDs and the elapsed runtime of all tasks having a priority greater then the input task.

c. The Queue Time Estimator then retrieves from the database the estimated run time of tasks with ID's retrieved in step (b). The run time of each task is estimated at the time of task submission and is stored in a separate database.

d. The elapsed run time of retrieved tasks is then subtracted from their estimated run time; this gives the remaining estimated run time for each task. The sum of the estimated remaining runtime of all tasks retrieved in step 2 is the estimated queue time for the input task.

### 6.3. File Transfer time Estimator

Since transferring files could be a time consuming operation, we need an estimator method that will tell the user how much time this file will take to transfer. For transfer time estimation, we first determine the bandwidth between the client and the Clarens server using iperf, and then using this bandwidth and the file size, we calculate the transfer time.

## 7 Performance Results

In order to test the Runtime Estimator, we used accounting data from the Paragon Supercomputer at the San Diego Supercomputing Center. This data was collected by Allen Downey in 1995. The accounting data had the following information recorded for each job: account name; login name; partition to which the job was allocated; the number of nodes for the job; the job type (batch or interactive); the job status (successful or not); the number of requested CPU hours; the name of the queue to which the job was allocated; the rate of charge for CPU hours and idle hours; and the task's duration in terms of when it was submitted, started, and completed.

The history consisted of 100 jobs and the runtime for 20 jobs was estimated; figure 5 shows the estimated and actual runtimes in each of the 20 cases:

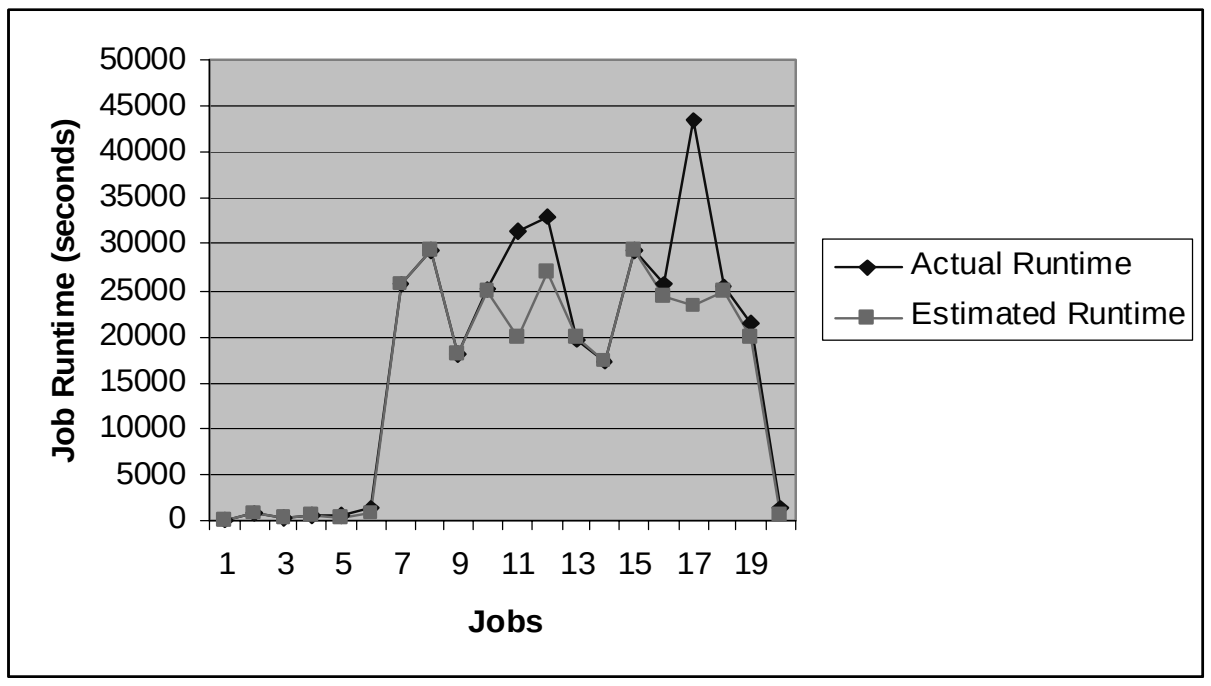

**Figure 5: Actual & Estimated Runtimes for 20 test cases**

For each of the twenty cases, we computed the error in estimation as:

Percentage Error = (Actual Runtime – Estimated Runtime)/Actual Runtime * 100 %

The percentage errors for the twenty cases were then used to compute the mean error of the runtime estimator. The mean error for the run time estimator comes out to be 13.53%, this was computed by dividing the sum of percentage errors in each of the twenty test cases by 20.

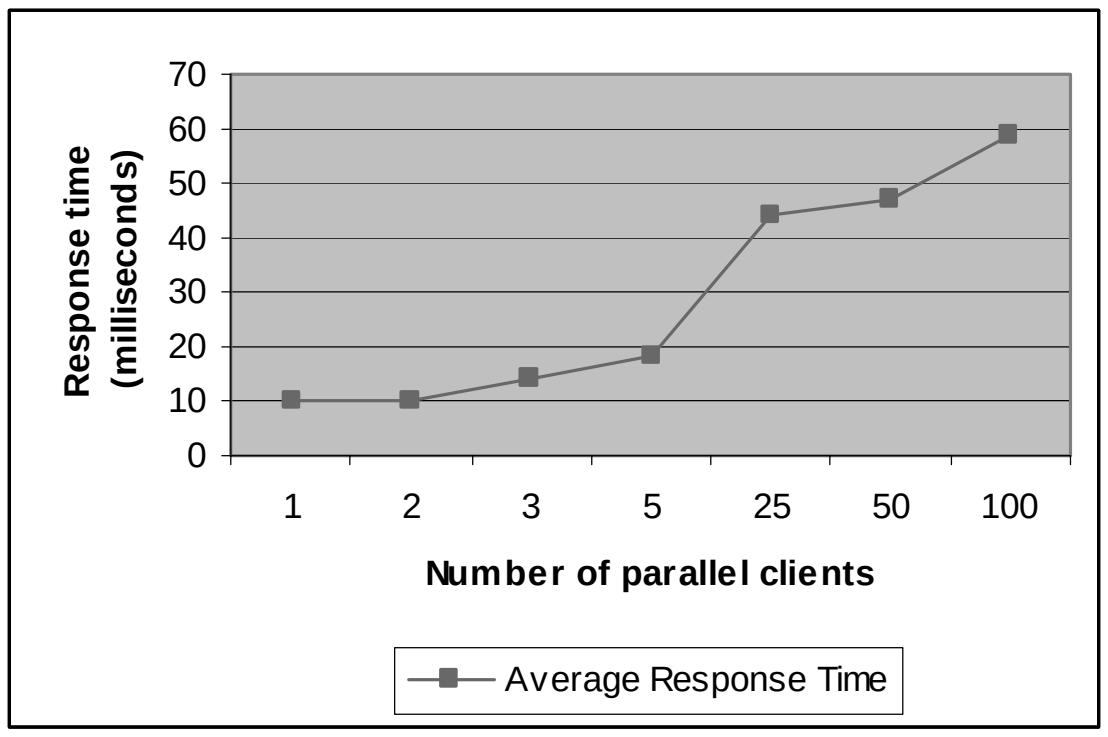

**Figure 6: Response times for queries to Job Monitoring Service**

We also carried out a number of tests to measure the performance of the Job Monitoring Service. We hosted the Job Monitoring Service on a Windows-XP based JClarens server. Several clients were initialized in parallel to call various methods of the service. Figure 6 shows the results in terms of the average time taken to fulfill a request when different numbers of clients tried to access the service concurrently.

The results show that the performance of the service scales well with increasing number of clients, which means that the service can handle a large number of clients as long as they do not exceed a certain limit.

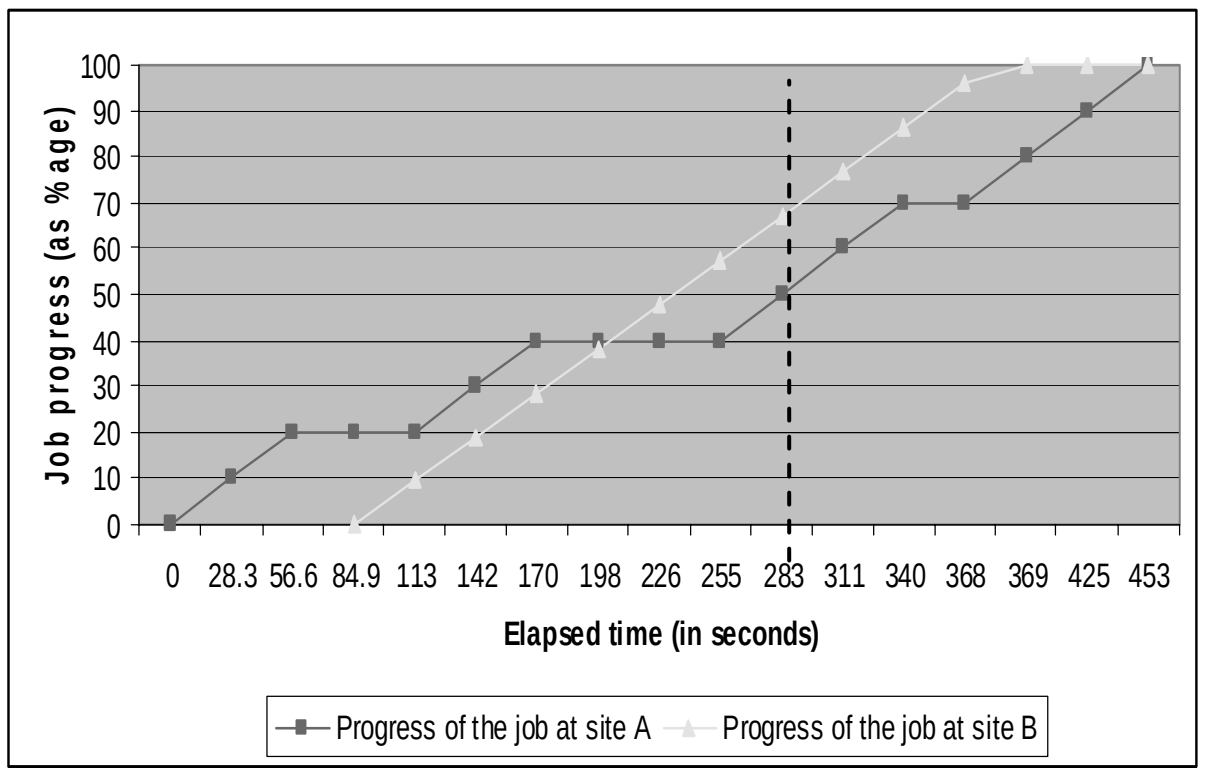

**Figure 7: Job Completion at different sites**

The improvement in performance of a grid that can be obtained by using the steering service is depicted in Figure 7, which shows the completion time of a job (a simple C++ program that calculates prime numbers over an input range) in different scenarios. The dashed line shows the estimated completion time of the job. Currently this estimate is calculated by running the job many times on different machines that have negligible CPU load. This estimate comes out to be 283 seconds. Hence, we make the assumption that the job requires 283 s to complete on a free CPU, i.e. if a CPU is free, the job will always complete in around 283 s. Moreover, Condor provides us the ability to see how much "wall-clock time" the job has accumulated while running on a node or Condor pool. Note that this "wall-clock" time does not include the time during which the job is idle and waiting for the CPU or other resources to free up.

We used this feature of Condor coupled with the above-mentioned assumption as a way to measure the progress of the job when it is running on a node with significant CPU load. Thus, if the job has accumulated 141 s of wall-clock time (as shown by Condor) when it is running on such a node, we assume that roughly 50% of the job is complete, even if the time elapsed since the job had been scheduled on that node is greater (because the job has had to remain idle in the queue, for instance).

Based on this data, we charted the progress of a job from 0 to 100% while it ran on a node A with significant CPU load. The purple line shows the job that is running on site A under significant CPU load. The steering service has monitored the progress of this job (using the job monitoring service) and has decided to move this job based on its slow execution rate (note that the user could have moved the job from site A to site B manually as well). This job is then rescheduled on some new site B, while the job was also allowed to continue running on site A for testing purposes. It is clear that after rescheduling, the job has completed much sooner than the job that was executing at site A (as indicated by the yellow line).

The job can be completed even quicker than 369 seconds if it is checkpoint-able and flocking is enabled between site A and Site B. A critical factor that affects the job completion time is the time at which the decision to move the job is taken. The quicker the decision is taken, the better the chance that it will complete quicker. Another important factor is the time taken to transfer the data files needed by the job. All of these factors must be taken into account when deciding whether a job should be transferred or allowed to run to completion.

We conclude that there are tangible benefits to be achieved by using the resource management services described in this paper. By allowing the steering service to periodically monitor the performance of the job (using the job monitoring service), to make dynamic estimates of the job completion time, and to reschedule the job when required, the productivity and throughput of the grid infrastructure is enhanced. Moreover, since the APIs of the steering service also enable users to get this information, advanced users can also make such rescheduling decisions when they think that the performance is insufficient. It might not be possible to get ideal performance from the system because it takes some time to detect the slow execution rate of a job. However, once this is detected, a rescheduling of the job based on updated monitoring information can dramatically reduce the actual execution time.

## 8. Related Work

MonALISA is perhaps the most relevant project currently being used to perform monitoring, steering, and optimization for the VRVS [11] reflector network. GMonitor [12] is a web portal that allows a user to monitor, control, and steer the execution of application jobs on global grids. G-Monitor does not provide an interface for applications to interact with the system. Falcon [13], from the Georgia Institute of Technology, was created to steer parallel programs through online monitoring, allowing Monitoring and Steering but not Optimization.

The Vase System [14] is another effort, which focuses on the steering of applications by human users. This project lacks autonomous decision-making and algorithmic program steering. NetLogger [15] is a Toolkit for Distributed System Performance Tuning and Debugging. This project provides methodology to get the detailed end-to-end application and system level monitoring.

Grid information systems such as Globus MDS [16] and R-GMA [17] provide some monitoring but they do not support all monitoring scenarios. To overcome these limitations, application monitoring solutions have been proposed such as NetLogger and OCM-G [18]. Application monitoring has been done in the CMS production tools IMPALA [19] and BOSS [20].

Previous efforts of application runtime estimation can be broadly classified in three categories:

1. Code analysis [21] techniques estimate execution by analyzing the source code of the task.
2. Analytic benchmarking/Code Profiling [22] defines a number of primitive code types. On each machine, benchmarks are obtained which determine the performance of the machine for each code type. The analytic benchmarking data and the code profiling data are then combined to produce an execution time estimate.
3. Statistical prediction [23] algorithms make predictions using past observations. This is the technique used by us for task runtime estimation.

Another notable effort is that of Sebastian Grinstein and John Huth [24] who devised a technique that aims at predicting the behavior of ATLAS applications. The technique they have used falls into the analytic benchmarking/code profiling category. Statistical runtime prediction techniques have also been used by Smith, Taylor and Foster [25] to predict the queue wait time of applications.

## 9. Conclusion

We have presented an ensemble of resource management services which working in tandem provide users with great flexibility in executing their jobs, monitoring the progress of their jobs, and getting improved performance from the Grid systems available to them. These services are crucial to the GAE because they ensure that users get maximum control over the execution of their jobs, and the locations from where the jobs access their required data.

## 10. ACKNOWLEDGEMENTS

This work is partly supported by the Department of Energy as part of the Particle Physics DataGrid project and by the National Science Foundation. Any opinions, findings, conclusions or recommendations expressed in this material are those of the authors and do not necessarily reflect the views of the Department of Energy or the National Science Foundation.